\documentclass[12pt]{article}
\usepackage{epsfig}
\newlength{\defbaselineskip}
\newcommand{\setlinespacing}[1]%
           {\setlength{\baselineskip}{#1 \defbaselineskip}}

\begin{document}

\title{The systematic study of the isotopic dependence of fusion dynamics for neutron and proton-rich nuclei using proximity formalism}

\author{O. N. Ghodsi, R. Gharaei\thanks{Email: r.gharaei@stu.umz.ac.ir}, and F. Lari \\
\\
{\small {\em  Sciences Faculty, Department of Physics, University of Mazandaran}}\\
{\small {\em P. O. Box 47415-416, Babolsar, Iran}}\\
}
\date{}
\maketitle

\begin{abstract}
\noindent The behaviors of barrier characteristics and fusion cross
sections are analyzed by changing neutron over wide range of
colliding systems. For this purpose, we have extended our previous
study (Eur. Phys. J. A \textbf{48}, 21 (2012), it is devoted to the
colliding systems with neutron-rich nuclei) to 125 isotopic systems
with condition of $0.5\leq N/Z \leq 1.6$ for their compound nuclei.
The AW 95, Bass 80, Denisov DP and Prox. 2010 potentials are used to
calculate the nuclear part of interacting potential. The obtained
results show that the trend of barrier heights $V_B$ and positions
$R_B$ as well as nuclear $V_N$ and Coulomb $V_C$ potentials (at
$R=R_B$) as a function of ($N/Z-1$) quantity are non-linear
(second-order) whereas the fusion cross sections follow a
linear-dependence.

\end{abstract}

\newpage
\setlinespacing{1.5}
{\noindent \bf{1. INTRODUCTION}}\\

The systematic study of the isotopic dependence of interacting
potential and fusion cross sections is one of the interesting
subjects in nuclear physics. It is carried out using different
theoretical models such as Skyrme energy density formalism,
Ng$\hat{o}$ and Ng$\hat{o}$, Christensen and Winther, Bass and
Denisov potentials [1-3]. Although, the choice of these potentials
was not made on merit to reproduce the experimental data. One can
divide the isotopic systems in previous studies into three series:
different collisions of Ca and Ni isotopes, with conditions of (i)
$1\leq N/Z \leq 2$ [1], (ii) $0.6\leq N/Z \leq 2$ [2] and (iii)
$0.5\leq N/Z \leq 2$ [3]. In these conditions $N$ and $Z$ are
neutron and proton numbers of compound nucleus. In all mentioned
investigations, the isotopic dependence of barrier heights $V_B$ and
positions $R_B$ as well as fusion cross sections $\sigma_{fus}$
versus $N/Z$ ratio, has been examined. The obtained results show
that the increasing trend of $R_B$ and $\sigma_{fus}$ and decreasing
trend of $V_B$ as a function of $N/Z$ ratio for fusion systems with
condition (i) are linear, see fig. 2 of Ref. [1]. On the other hand,
in different collisions of Ca and Ni isotopes which have condition
of $0.6\leq N/Z \leq 2$ (or $0.5\leq N/Z \leq 2$), it is shown that
this behavior is non-linear (second-order) for $R_B$ and $V_B$
whereas the fusion cross sections follow a linear-dependence (see
corresponding figures of Refs. [2,3]). As a result, one expects that
with increase of neutron and its effect on the attractive force, the
Coulomb barrier heights decrease. Since the fusion probability is
directly dependent on these parameters, therefore it is predictable
that the fusion cross sections increase with addition of neutron in
different isotopic systems. As an important issue, it can be noted
that the experimental data have only been reported for fusion
reactions with nuclei which are near the stability line ($N=Z$).
Therefore, one can compare the measured and calculated values of
fusion cross sections for limited numbers of reactions. However,
using the proposed semi-empirical approaches, such as Ref. [4], the
experimental data are well described.

In a recent study, we have analyzed the isotopic dependence of
fusion cross sections and barrier characteristics for 50 fusion
reactions with condition of $1\leq N/Z \leq 1.6$ [5]. For this
purpose, we have selected the fusion reactions which the C, O, Mg,
Si, S, Ca, Ar, Ti and Ni isotopes are as their participant nuclei.
In this study, the nuclear part of interacting potential has been
calculated using four versions of proximity model, namely AW 95 [6],
Bass 80 [7,8], Denisov DP [9] and Prox. 2010 [10] potentials. It is
shown that these models have good agreement with experimental data
[10-12]. Our obtained results, like Ref. [1], confirmed the linear
trend of $R_B$, $V_B$ and $\sigma_{fus}$ versus $N/Z$ ratio for
fusion reactions with $1\leq N/Z \leq 1.6$.

In present work, we are going to extend our studies to both proton
and neutron-rich systems. For this purpose, we have chosen 125
fusion reactions so that all colliding pairs are assumed to be
spherical and the $N/Z$ ratio of compound nuclei, which are formed
during fusion process, to be in the range $0.5\leq N/Z \leq 1.6$.
For neutron-deficient systems, we have only intended to investigate
the influence of changing neutron on the input potential channel and
fusion probabilities. The lightest ($^{10}$C) and heaviest
($^{54}$Ni) proton-rich nuclei have the half-life $T_{1/2}=19.30$ s
and $T_{1/2}=104.00$ ms, respectively. The proximity formalism and
Wong model [13] are employed to calculate the nuclear potentials and
fusion cross sections, respectively.

In summery, our motivations in this work are: i)  There is no
systematic study on the isotopic dependence of fusion cross sections
based on the AW 95, Bass 80, Denisov and Prox. 2010 potentials in
the range of $N/Z\leq1$. The applied models in previous studies
[1-3] were not made on merit to reproduce the experimental data
whereas our selected models are able to reproduce experimental data
within $\% 10$, on the average, see Refs. [10-13] for details. ii)
The different colliding systems in previous works [1-3] are only
included Ca and Ni isotopes. Whereas, we have used the C, O, Mg, Si,
S, Ca, Ar, Ti and Ni nuclei as colliding pairs which have been taken
from proton-rich region of periodic table. Our selections can be
more appropriate for better understanding of the isotopic dependence
of $R_B$, $V_B$ and $\sigma_{fus}$ versus $N/Z$ ratio. iii) The
isotopic systems which are used in previous studies such as Refs
[1-3] lying far from the stability line (N=Z) and in these regions
the experimental data have not been reported. Whereas, using present
study, one can analyze the calculated results and compare them with
corresponding experimental data. iv) Such studies can be very useful
to predict the properties of new and superheavy elements which are
produced in fusion process and are not available at present.

The study of neutron rich nuclei is also reported at heavy-ion
collisions with intermediate energies. The effects of isospin degree
of freedom in collective and elliptic flow have been studied, for
example, in Refs. [14-17]. Recently, using the dynamic approach
based on the macroscopic models, the isospin effects have been
examined for $^{40}$Ca+$^{90,96}$Zr, $^{48}$Ca+$^{90}$Zr fusion
reactions [18]. The obtained results reveal that the dynamic effects
decrease barrier height and thickness.

The paper is organized as follows: in sections 2, we discuss about
the nuclear part of the total interaction potential as well as the
employed models for calculation it. The analysis of isotopic
dependence of barrier characteristics and fusion cross sections in
different ranges of $N/Z$ ratio have been carried out in Secs. 3 and
4. Section 5 is devoted to some concluding remarks.
\\
\\

\noindent{\bf {2. THE NUCLEAR POTENTIAL OF INTERACTING SYSTEMS}}\\

In general, with assumption that the participant nuclei in fusion
reaction to be in the s-wave state ($\ell=0$), the total potential
can be defined as the sum of two parts which are caused by
electrostatic (Coulomb-repulsion) and strong (nuclear-attraction)
interactions. In recent years, many theoretical models have been
introduced to parameterize the last interactions. The proximity
formalism is one of the useful models for calculating of nuclear
potential. The various versions of this formalism are introduced in
[10-12]. These studies have been carried out on the many different
systems. The obtained results show that all introduced models
determine the fusion barrier heights with accuracy $\pm10\%$, on the
average. Among various versions, our selected potentials, namely AW
95, Bass 80, Denisov DP and Prox. 2010, reproduce the best results
for potential and fusion cross sections. These models are briefly
explained in the following.

Aage winther proposed a nuclear potential which is parameterized
based on the Woods-Saxon form [6],

\begin{equation} \label{1}
V^{AW 95}_{N}(r)=-\frac{V_{0}}{1+ \textmd{exp} (\frac{r-R_0}{a})},
\end{equation}
here, the $V_0$, $R_0$ and $a$ parameters are defined as,

\begin{equation} \label{2}
V_{0}=16\pi\frac{R_{1}R_{2}}{R_1+R_2}\gamma a,
\end{equation}

\begin{equation} \label{3}
R_0=R_1+R_2,
\end{equation}
and
\begin{equation} \label{4}
a=\bigg[\frac{1}{1.17(1+0.53(A_1^{-1/3}+A_2^{-1/3}))}\bigg].
\end{equation}
The $R_i$ and $\gamma$ parameters in equations (2) and (3)
respectively stand for the radius of target/projectile and surface
energy coefficient and can be written as,

\begin{equation} \label{5}
R_i=1.2A_i^{1/3}-0.09 \qquad (i=1, 2),
\end{equation}

\begin{equation} \label{6}
\gamma=0.95\bigg[1-1.8\bigg(\frac{N_p-Z_p}{A_p}\bigg)\bigg(\frac{N_t-Z_t}{A_t}\bigg)\bigg].
\end{equation}
where $A_{p(t)}$, $Z_{p(t)}$ and $N_{p(t)}$ are characteristics of
target and projectile.

According to the proximity theorem [19], the nuclear potentials
which are based on the models such as Bass 80, Denisov and Porx.
2010 define as the product of a geometrical factor and a universal
function which are respectively dependent on the mean curvature of
the interaction surface and the separation distance. Therefore, one
can use the below equations to calculate the nuclear potential based
on the selected models,

\begin{equation} \label{7}
V^{Bass 80}_{N}(r)=-\frac{R_{1}R_{2}}{R_1+R_2}\Phi(s=r-R_1-R_2),
\end{equation}

\begin{eqnarray} \label{8}
V^{Denisov DP}_{N}(r)=-1.989843\frac{R_{1}R_{2}}{R_1+R_2}\Phi(s=r-R_1-R_2-2.65)\nonumber\\
\times\bigg[1+0.003525139(\frac{A_1}{A_2}+\frac{A_2}{A_1})^{{3}/{2}}-0.4113263(I_1+I_2)\bigg],
\end{eqnarray}

\begin{equation} \label{9}
V^{Prox.2010}_{N}(r)=4\pi\gamma\frac{R_{1}R_{2}}{R_1+R_2}\Phi(s=r-C_1-C_2).
\end{equation}
In these relations the radius parameter $R_i$ can be written as
follows,

\begin{equation} \label{10}
R_{i}^{Bass 80}=R_{s}\bigg(1-\frac{0.98}{R^{2}_{s}}\bigg) \qquad
(i=1,2),
\end{equation}

\begin{equation} \label{11}
R_{i}^{Denisov DP}=1.2332A_i^{1/3}(1+2.348443/A_i-0.151541A_{si})
\qquad (i=1,2),
\end{equation}
where the sharp radius $R_{s}$ in Eq. (10) is given as $R_s=1.28
A^{1/3} - 0.76 + 0.8A^{-1/3}$. The $R_i$ parameter in Prox. 2010 is
similar to AW 95 model, Eq. (5). The universal function $\Phi(s)$ is
respectively defined by Eqs. (20), (42) and (6) of Ref. [11] for
Bass 80, Denisov DP and Prox. 2010 potentials.
\\
\\
\noindent{\bf {3. ISOTOPIC ANALYSIS OF TOTAL POTENTIALS}}\\

By adding the Coulomb part to our selected nuclear potentials, which
are introduced in previous sections, one can calculate total
interaction potential for different fusion systems by following
simple expression,

\begin{equation} \label{10}
V_{tot}(r)=V_{N}(r)+V_{C}(r)= V_{N}(r)+\frac{Z_1 Z_2 e^2}{r},
\end{equation}
where $Z_1$ and $Z_2$ are atomic numbers of interaction nuclei. The
effects of addition/removal of neutron on the various potentials are
shown in Fig. 1. In this figure, the nuclear $V_{N}(r)$, Coulomb
$V_{C}(r)$ and total $V_{T}(r)$ potentials are separately plotted
for $^{A_{1}}$Ni+$^{A_{2}}$Ni isotopic systems, which are consist of
proton-rich ($^{48}$Ni+$^{48}$Ni, $^{50}$Ni+$^{50}$Ni and
$^{54}$Ni+$^{54}$Ni), neutron-rich ($^{58}$Ni+$^{58}$Ni and
$^{64}$Ni+$^{64}$Ni) as well as symmetric ($^{56}$Ni+$^{56}$Ni)
participant nuclei. Using the results shown in figure 1, the
influence of neutron-excess on the shape of considered potentials is
quite evident.

The exact values of the barrier height $V_B^{theor}$ and barrier
position $R_B^{theor}$ have been extracted using

\begin{equation} \label{11}
{\bigg(\frac{dV_{tot}(r)}{dr}\bigg)}_{r=R^{theor}_B}=0\qquad;\qquad{\bigg(\frac{d^2V_{tot}(r)}{dr^2}\bigg)}_{r=R^{theor}_B}\leq{0}.
\end{equation}
In the beginning, we have calculated the barrier heights and
positions for different fusion systems using four-type of proximity
potentials, namely AW 95, Bass 80, Denisov DP and Prox. 2010. The
obtained results of these calculations are listed in Table 1. It can
be seen that with addition of neutron in the interacting systems,
the barrier heights and positions, respectively, decrease and
increase. Moreover, using definition of Coulomb potential
$V_C(r)={Z_1 Z_2 e^2}/{r}$, we expect that by increasing $R_B$ in
any isotopic system the values of $V_C(r=R_B)$ reduce.

To get a better comparison between experimental and theoretical
values of barrier heights and positions, we have displayed the
$V_{B}^{theor}$ vs $V_{B}^{exp}$ and the $R_{B}^{theor}$ vs
$R_{B}^{exp}$, see Figs. 2 and 3. It is shown that the calculated
values of $V_{B}^{theor}$ based on the considered models have good
compatibility with experimental data. On the other hand, according
to Fig. 3, one can't find a regular behavior in predictions of
barrier positions. This may be caused by the large uncertainties of
these values.
\\
\\
\noindent{\bf {3.1 The isotopic dependence of barrier heights $V_B$ and positions $R_B$}}\\

The percentage difference of barrier characteristics, i.e. $\Delta
R_B(\%)$ and $\Delta V_B(\%)$, are defined as follows to study the
isotopic dependence of barrier heights and positions,

\begin{equation} \label{12}
\Delta{R_B(\%)}=\frac{R_B-R_B^0}{R_B^0}\times100,
\end{equation}

\begin{equation} \label{13}
\Delta{V_B(\%)}=\frac{V_B-V_B^0}{V_B^0}\times100.
\end{equation}
Above, $R_B^0$ and $V_B^0$ are barrier characteristics of $N=Z$
case. Indeed, using the proposed producer of Refs. [1-3], our
criterion to analyze the trend of $R_B$ and $V_B$ versus $N/Z$ ratio
in each set of colliding systems is the symmetric reaction of that
set. Using the straight-line interpolation between know values, we
have estimated the barrier characteristics ($R_B$ and $V_B$) for
symmetric reactions ($N=Z$) that these values aren't available for
them. The obtained results for Eqs. (14) and (15) based on the
selected proximity potentials have been plotted in Fig. 4. It is
clear that with addition of neutron, the values of $R_B$ and $V_B$
respectively increase and decrease. Moreover, figure 4 shows the
regular behaviors for barrier heights and positions. One can analyze
these behaviors using the following ranges of $N/Z$ ratio.
\\
\\
\noindent{\bf {3.1.1 The ranges of $0.5\leq N/Z \leq 1$ and $1\leq N/Z \leq 1.6$}}\\

The values of barrier heights and positions based on the considered
proximity potentials follow a linear-dependence for either proton
($0.5\leq N/Z \leq 1$) or neutron-rich ($1\leq N/Z \leq 1.6$)
systems (see Fig. 4). One can parameterize the percentage difference
of $R_B$ and $V_B$, which are calculated by Eqs. (14) and (15),
using the below forms in two mentioned regions, namely for
$N/Z\leq1$,

\begin{equation} \label{14}
\Delta{R_B(\%)}=\alpha_{1}\bigg(\frac{N}{Z}-1\bigg); \qquad
\Delta{V_B(\%)}=\alpha_{2}\bigg(\frac{N}{Z}-1\bigg),
\end{equation}
and for $N/Z\geq1$,

\begin{equation} \label{15}
\Delta{R_B(\%)}=\alpha_{1}^{'}\bigg(\frac{N}{Z}-1\bigg),\qquad
\Delta{V_B(\%)}=\alpha_{2}^{'}\bigg(\frac{N}{Z}-1\bigg),
\end{equation}
where the values of the constants $\alpha_i$ and $\alpha^{'}_i$, for
i=1, 2, have been listed in Table 2.
\\
\\
\noindent{\bf {3.1.2 The range of $0.5\leq N/Z \leq 1.6$}}\\

In whole region of $0.5\leq N/Z \leq 1.6$, the behavior of heights
and positions of the barrier are non-linear and the percentage
difference of these values can be parameterized by a second-order
form (see figure 4),

\begin{equation} \label{16}
\Delta{R_B(\%)}=\beta_{1}\bigg(\frac{N}{Z}-1\bigg)+\beta_{2}{\bigg(\frac{N}{Z}-1\bigg)}^2;
\qquad
\Delta{V_B(\%)}=\beta_{3}\bigg(\frac{N}{Z}-1\bigg)+\beta_{4}{\bigg(\frac{N}{Z}-1\bigg)}^2
\end{equation}
where the values of the coefficients $\beta_{i}$ have been listed in
Table 3. In Fig. 4, we have also plotted the results of two
theoretical models [3,20], for example. It is clear that the
calculated values of $\Delta{R_B(\%)}$ and $\Delta{V_B(\%)}$ for
these models are consistent with our predictions. On the other hand,
the results shown in Fig. 4 confirm the trend of $R_B$ and $V_B$
which are reported in Refs. [2,3].
\\
\\
\noindent{\bf {3.2 The isotopic dependence of nuclear $V_N$ and Coulomb $V_C$ potentials}}\\

In addition to $R_B$ and $V_B$, we have interested to analyze the
isotopic dependence of nuclear and Coulomb potentials (at $r=R_B$)
by changing neutron. For this aim, one should calculate the values
of $\Delta V_N(\%)$ and $\Delta V_C(\%)$ using the following
relations,

\begin{equation} \label{17}
\Delta{V_N(\%)}=\frac{V_N-V_N^0}{V_N^0}\times100.
\end{equation}

\begin{equation} \label{18}
\Delta{V_C(\%)}=\frac{V_C-V_C^0}{V_C^0}\times100.
\end{equation}
where $V_N^0$ and $V_C^0$ are the values of nuclear and Coulomb
potentials for symmetric reaction, see Fig. 5. Similar previous
calculations of $R_B$ and $V_B$ and using the above suggested
manner, one can predict the values of $V_N(r=R_B)$ and $V_C(r=R_B)$
for symmetric reactions that these values aren't available for them.
Because of increasing neutron, it is predictable that the values of
nuclear and Coulomb potentials for different isotopic systems
increase and decrease, respectively. The regular behaviors of $V_N$
and $V_C$ at $r=R_B$ are examined in following ranges.
\\
\\
\noindent{\bf {3.2.1 The ranges of $0.5\leq N/Z \leq 1$ and $1\leq N/Z \leq 1.6$}}\\

The values of Coulomb and nuclear potentials based on the AW 95,
Bass 80, Denisov DP and Prox. 2010 potentials follow a
linear-dependence for either proton ($0.5\leq N/Z \leq 1$) or
neutron-rich ($1\leq N/Z \leq 1.6$) systems (see Fig. 5). The
percentage differences of $V_C$ and $V_N$ are parameterized as
following forms for $N/Z\leq1$,

\begin{equation} \label{19}
\Delta{V_C(\%)}=\alpha_{3}\bigg(\frac{N}{Z}-1\bigg),\qquad
\Delta{V_N(\%)}=\alpha_{4}\bigg(\frac{N}{Z}-1\bigg),
\end{equation}
and for $N/Z\geq1$,

\begin{equation} \label{20}
\Delta{V_C(\%)}=\alpha_{3}^{'}\bigg(\frac{N}{Z}-1\bigg) ;\qquad
\Delta{V_N(\%)}=\alpha_{4}^{'}\bigg(\frac{N}{Z}-1\bigg),
\end{equation}
where the values of constants $\alpha_i$ and $\alpha^{'}_i$, for
i=3, 4, have been listed in Table 2.
\\
\\
\noindent{\bf {3.2.2 The range of $0.5\leq N/Z \leq 1.6$}}\\

In range of $0.5\leq N/Z \leq 1.6$, we have found a non-linear
regular behavior for both nuclear and Coulomb potentials. One can
formulate these trends as,
\begin{equation} \label{21}
\Delta{V_C(\%)}=\beta^{'}_{1}\bigg(\frac{N}{Z}-1\bigg)+\beta^{'}_{2}{\bigg(\frac{N}{Z}-1\bigg)}^2;
\qquad
\Delta{V_N(\%)}=\beta^{'}_{3}\bigg(\frac{N}{Z}-1\bigg)+\beta^{'}_{4}{\bigg(\frac{N}{Z}-1\bigg)}^2,
\end{equation}
where the values of coefficients $\beta^{'}_{i}$ have been listed in
Table 3.
\\
\\
\noindent{\bf {4. ISOTOPIC ANALYSIS OF FUSION CROSS SECTIONS}}\\

In present study, one dimensional penetration model [21,22] is used
to study the fusion cross section, $\sigma_{fus}$. In this
formalism, the $\sigma_{fus}$ defines as the summation on the
quantum-mechanical transmission probability through the potential
barrier for a specified angular momentum $l$ and center-of-mass
energy, namely $T_{l}(E_{c.m.})$,

\begin{equation} \label{22}
\sigma_{fus}=\frac{\pi \hbar^{2}}{2\mu
E_{c.m.}}\sum_{l=0}^{l_{max}}(2l+1)T_{l}(E_{c.m.}),
\end{equation}
where in this relation $\mu$ is the reduced mass of interacting
system. With assumption that $E_{c.m.}\gg V_{B}$, the Eq. (24)
reduces to the well-known sharp cutoff formula,

\begin{equation} \label{23}
\sigma_{fus}=10\pi R^{2}_{B}(1-\frac{V_{B}}{E_{c.m.}}).
\end{equation}
Among introduced potentials in Refs. [10-12], we have selected the
models that have the best agreement with fusion data. To get a
better comparison, the fusion cross sections for
$^{A_1}$Ca+$^{A_2}$Ti system, namely $^{40}$Ca+$^{46}$Ti,
$^{40}$Ca+$^{48}$Ti and $^{40}$Ca+$^{50}$Ti fusion reactions, are
calculated by sharp cutoff formula, Eq. (25). For this purpose, we
have used the obtained results of $R_B$ and $V_B$ based on the AW
95, Bass 80, Denisov DP and Prox. 2010 potentials (see Fig. 6). In
this figure, the corresponding experimental data for considered
reactions are taken from [23]. It is shown that the predicted values
for fusion cross sections are consistent with experimental data
particularly at above barrier energies.

To systematic study of fusion cross section in isotopic systems, we
have defined the percentage difference of this quantity, namely
$\Delta \sigma_{fus}(\%)$, as following form,

\begin{equation} \label{24}
\Delta{\sigma_{fus}(\%)}=\frac{\sigma_{fus}(E_{c.m.}^0)-\sigma^{0}_{fus}(E_{c.m.}^0)}{\sigma^{0}_{fus}(E_{c.m.}^0)}\times100,
\end{equation}
where the $\sigma_{fus}(E_{c.m.}^0)$ is fusion cross section for
reaction with $N=Z$ condition. We have computed the $\Delta
\sigma_{fus}(\%)$ at some above barrier energies such as
$E_{c.m.}=1.125V_{B}^{0}$ and $E_{c.m.}=1.375V_{B}^{0}$. The
calculated results have been shown in Fig. (7). It is clear that by
increasing neutron in interacting systems and decreasing barrier
height, one expects that fusion cross sections enhance. This
behavior is quite obvious in Fig. (7). In contrast the obtained
results for barrier characteristics , the relationship between
variations of fusion cross sections $\Delta \sigma_{fus}(\%)$ and
increasing of ($N/Z-1$) quantity is linear. This subject is accurate
for all neutron and proton-rich systems ($0.5\leq N/Z \leq 1.6$). We
have parameterized this linear trend of fusion cross sections as
following form,

\begin{equation} \label{25}
\Delta{\sigma_{fus}(\%)}=\gamma\bigg(\frac{N}{Z}-1\bigg),
\end{equation}
where the values of constant coefficient $\gamma$ have been listed
in Table 4 for various potentials and energies.
\\
\\
\noindent{\bf {5. CONCLUSIONS}}\\

In this paper, using the systematic study on the large range of
colliding pairs C, O, Mg, Si, S, S, Ca, Ar, Ti and Ni with $84\leq
Z_1Z_2 \leq 784$, we have analyzed the isotopic dependence of
different parameters of interacting potentials and fusion cross
sections. For calculating of these values, we have respectively used
four confirmed versions of proximity formalism and Wong model. Our
obtained results for three considered regions of $N/Z$ ratio, i.e.
$0.5\leq N/Z \leq 1$, $1\leq N/Z \leq 1.6$ and $0.5\leq N/Z \leq
1.6$ are: a) For fusion systems with condition of $0.5\leq N/Z \leq
1$, the variations trend of barrier characteristics, $R_B$,
$V_C(R=R_B)$, $V_N(R=R_B)$ and $V_B$, with respect to corresponding
symmetric reaction ($N=Z$), follow a linear dependence as a function
of (N/Z-1). b) For colliding systems with condition of $1\leq N/Z
\leq 1.6$, the above quantities can be parameterized as linear. c)
In whole range $0.5\leq N/Z \leq 1.6$, the values of $\Delta R_B$,
$\Delta V_C$, $\Delta V_N$, and $\Delta V_B$ have different trends.
In other words, these values follow a non-linear second order
behavior with addition/removal neutron.

As a common property, one can point out the values of $R_B$,
$V_N(R=R_B)$ as well as $V_B$, $V_C(R=R_B)$, respectively, increase
and decrease with increasing of neutron. On the other hand, the
fusion cross sections follow a linear dependence for all considered
isotopic systems.
\\
\\

\newpage
\noindent{\bf {FIGURE CAPTIONS}}\\
\\
Fig. 1. The various components of interacting potential, i.e.
nuclear $V_{N}$, Coulomb $V_{C}$ and total $V_{T}$ potentials,
versus inter-nuclear distance r (in fm) for $^{A_1}$Ni+$^{A_2}$Ni
isotopic system. These calculations are based on the Prox. 2010 and
Denisov DP potentials, for example.
\\
\\
Fig. 2. The obtained theoretical barrier heights $V_B^{theor}$ (in
MeV) as a function of corresponding experimental data $V_B^{exp}$
(in MeV) [24-40] based on the AW 95, Bass 80, Denisov DP and Prox.
2010 potentials.
\\
\\
Fig. 3. The obtained theoretical barrier positions $R_B^{theor}$ (in
MeV) as a function of corresponding experimental data $R_B^{exp}$
(in MeV) [24-40] based on the AW 95, Bass 80, Denisov DP and Prox.
2010 potentials.
\\
\\
Fig. 4. The variations trend of $R_B$ ($\Delta R_B (\%)$, left
panels) and $V_B$ ($\Delta V_B (\%)$, right panels) as a function of
($N/Z-1$) based on the selected potentials AW 95, Bass 80, Denisov
DP and Prox. 2010. The dash and short-dotted lines are respectively
used to extract the linear-dependence of $\Delta R_B (\%)$ and
$\Delta V_B (\%)$ values in $0.5\leq N/Z \leq 1$ and $1\leq N/Z \leq
1.6$ regions. The solid lines are caused by the non-liner (second
order) fitting to the calculated values in whole range of $0.5\leq
N/Z \leq 1.6$.
\\
\\
Fig. 5. The variations trend of $V_N$ ($\Delta V_N (\%)$, left
panels) and $V_C$ ($\Delta V_C (\%)$, right panels) as a function of
($N/Z-1$) based on the selected potentials AW 95, Bass 80, Denisov
DP and Prox. 2010. The dash and short-dotted lines are respectively
used to extract the linear-dependence of $\Delta R_B (\%)$ and
$\Delta V_B (\%)$ values in $0.5\leq N/Z \leq 1$ and $1\leq N/Z \leq
1.6$ regions. The solid lines are caused by the non-liner (second
order) fitting to the calculated values in whole range of $0.5\leq
N/Z \leq 1.6$.
\\
\\
Fig. 6 The comparison of theoretical, Eq. (25), and experimental
data [23] for fusion cross sections based on the various versions of
proximity formalism, namely AW 95, Bass 80, Denisov DP and Prox.
2010 potentials. These calculations have been carried out for
$^{40}$Ca+$^{46}$Ti, $^{40}$Ca+$^{48}$Ti and $^{40}$Ca+$^{50}$Ti
fusion reactions.
\\
\\
Fig. 7. The percentage difference of fusion cross sections
$\Delta\sigma_{fus} (\%)$ as a function of ($N/Z-1$) for different
colliding systems. These values are calculated for two instance of
above barrier energies, namely $E_{c.m.}=1.125V_B^0$ (left panels)
and $E_{c.m.}=1.375V_B^0$ (right panels), which are based on the (a)
AW 95, (b) Bass 80, (c) Denisov DP, and (d) Prox. 2010 potentials.
The linear-behavior of $\sigma_{fus}$ have been parameterized by
solid-lines (Eq. (27)).

\newpage
\noindent{\bf {TABLE CAPTIONS}}

Table 1. The obtained values for barrier positions $R_B$ and
heights $V_B$ based on the AW 95, Bass 80, Denisov DP and Prox. 2010 potentials
for neutron-deficient and -rich systems.

\begin{center}
\begin{tabular}{c c c c c c c c c c}
  \hline
  \hline
  Reaction & N/Z& $R_B^a$   &  $V_B^a$   &  $R_B^b$   &  $V_B^b$ & $R_B^c$   &  $V_B^c$ & $R_B^d$   &  $V_B^d$  \\
  &  &  (fm) & (MeV)& (fm)& (MeV) & (fm) & (MeV) & (fm) & (MeV) \\
  \cline{1-10}
  $^{10}$C+$^{22}$Si     &   0.6     &   7.76   &   14.35  &  7.78   &   11.14 &   7.40   &   14.76 &   7.34  &   14.79  \\
  $^{12}$C+$^{22}$Si     &   0.7     &   8.08   &   13.82  &  8.04   &   13.73 &   7.81   &   14.06 &   7.75  &   14.09  \\
  $^{12}$C+$^{24}$Si     &   0.8     &   8.22   &   13.61  &  8.19   &   13.51 &   8.02   &   13.71 &   7.98  &   13.73  \\
  $^{12}$C+$^{26}$Si     &   0.9     &   8.35   &   13.41  &  8.32   &   13.32 &   8.20   &   13.41 &   8.15  &   13.47  \\
  $^{12}$C+$^{28}$Si     &   1       &   8.47   &   13.23  &  8.45   &   13.14 &   8.36   &   13.15 &   8.29  &   13.26  \\
  $^{12}$C+$^{29}$Si     &   1.05    &   8.53   &   13.15  &  8.51   &   13.05 &   8.43   &   13.04  &   8.35 &   13.18  \\
  $^{12}$C+$^{30}$Si     &   1.1     &   8.58   &   13.07  &  8.57   &   12.97 &   8.50   &   12.92  &   8.41 &   13.10  \\
  \\
  $^{12}$O+$^{20}$Mg      &   0.6 &   7.73  &   16.43  &  7.76   &   16.18 & 7.24   &  17.09&   7.33   &   16.94 \\
  $^{12}$O+$^{22}$Mg      &   0.7 &   7.94  &   16.05  &  7.92   &   15.89 & 7.50   &  16.62&   7.63   &   16.33 \\
  $^{12}$O+$^{24}$Mg      &   0.8 &   8.12  &   15.72  &  8.07   &   15.64 & 7.70   &  16.23&   7.85   &   15.92 \\
  $^{14}$O+$^{24}$Mg      &   0.9 &   8.33  &   15.34  &  8.30   &   15.26 & 8.08   &  15.53&   8.12   &   15.45 \\
  $^{16}$O+$^{24}$Mg      &   1   &   8.52  &   15.02  &  8.50   &   14.93 & 8.37   &  15.01&   8.34   &   15.08 \\
  $^{16}$O+$^{26}$Mg      &  1.1  &   8.65  &   14.81  &  8.65   &   14.72 & 8.53   &  14.72&   8.47   &   14.88 \\
  $^{18}$O+$^{24}$Mg      &   1.1 &   8.70  &   14.74  &    8.68 & 14.65   & 8.60   &  14.60&   8.52   &   14.80 \\
  \\
  $^{12}$O+$^{22}$Si     &   0.545   &  7.72   &   19.18  &  7.77   &   18.83 &7.18   &   20.04 &  7.21    &   20.03  \\
  $^{14}$O+$^{22}$Si     &   0.636   &  8.02   &   18.53  &  8.00   &   18.35 &7.59   &   19.12 &  7.64    &   19.04  \\
  $^{16}$O+$^{22}$Si     &   0.727   &  8.27   &   18.02  &  8.21   &   17.94 &7.92   &   18.45 &  7.95    &   18.36  \\
  $^{16}$O+$^{24}$Si     &   0.818   &  8.41   &   17.74  &  8.36   &   17.67 &8.13   &   18.00 &  8.16    &   17.94  \\
  $^{16}$O+$^{26}$Si     &   0.909   &  8.54   &   17.49  &  8.50   &   17.41 &8.32   &   17.62 &  8.33    &   17.62  \\
  $^{16}$O+$^{28}$Si     &   1       &  8.66   &   17.26  &  8.63   &   17.18 &8.48   &   17.29 &  8.46    &   17.36  \\
  $^{16}$O+$^{29}$Si     &   1.045   &  8.72   &   17.15  &  8.69   &   17.07 &8.56   &   17.14 &  8.52    &   17.25  \\
  $^{16}$O+$^{30}$Si     &   1.090   &  8.77   &   17.05  &  8.75   &   16.96 &8.63   &   17.00 &  8.58    &   17.15  \\
  $^{18}$O+$^{28}$Si     &   1.090   &  8.83   &   16.94  &  8.81   &   16.86 &8.72   &   16.82 &  8.64    &   17.03  \\
  \\
  $^{20}$Mg+$^{30}$S     &   0.786  &   8.66  &   29.50  &   8.60  &   29.42 & 8.25   &   30.35 &   8.40      &   29.97 \\
  $^{22}$Mg+$^{30}$S     &   0.857  &   8.82  &   29.01  &   8.76  &   28.95 & 8.48   &   29.59 &   8.59      &   29.37 \\
  $^{20}$Mg+$^{32}$S     &   0.857  &   8.79  &   29.10  &   8.72  &   29.07 & 8.40   &   29.85 &   8.55      &   29.50 \\
  $^{22}$Mg+$^{32}$S     &   0.928  &   8.94  &   28.65  &   8.88  &   28.60 & 8.63   &   29.12 &   8.73      &   28.96 \\
  $^{24}$Mg+$^{32}$S     &   1      &   9.08  &   28.24  &   9.03  &   28.18 & 8.84   &   28.51 &   8.88      &   28.50 \\
  $^{24}$Mg+$^{34}$S     &   1.071  &   9.18  &   27.94  &  9.14   &   27.88 &8.98   &   28.10 &   8.99      &   28.20  \\
  $^{26}$Mg+$^{32}$S     &   1.071  &   9.21  &   27.87  &  9.17   &   27.80 &9.02   &   27.99 &   9.01      &   28.12  \\
  $^{26}$Mg+$^{34}$S     &   1.143  &   9.31  &   27.60  &  9.28   &   27.50 &9.16   &   27.60 &   9.11      &   27.84  \\
  \hline
\end{tabular}
\end{center}

(a) Based on the AW 95 potential (b) Based on the Bass 80 potential
\\
(c) Based on the Denisov DP potential (d) Based on the Prox. 2010 potential

\newpage
Table 1. (Continued)

\begin{center}
\begin{tabular}{c c c c c c c c c c}
  \hline
  \hline
  Reaction & N/Z& $R_B^a$   &  $V_B^a$   &  $R_B^b$   &  $V_B^b$ & $R_B^c$   &  $V_B^c$ & $R_B^d$   &  $V_B^d$  \\
  &  &  (fm) & (MeV)& (fm)& (MeV) & (fm) & (MeV) & (fm) & (MeV) \\
  \hline
  \cline{1-10}
  $^{22}$Si+$^{22}$Si    &   0.571  &   8.18   &   31.70 &   8.19   &   31.35 &  7.68  &  33.09   &   7.73  &   32.98 \\
  $^{22}$Si+$^{24}$Si    &   0.643  &   8.37   &   31.06 &   8.34   &   30.85 &  7.90  &  32.26   &   8.01  &   31.94 \\
  $^{24}$Si+$^{24}$Si    &   0.714  &   8.54   &   30.50 &   8.49   &   30.37 &  8.12  &  31.47   &   8.24  &   31.12 \\
  $^{24}$Si+$^{26}$Si    &   0.786  &   8.69   &   30.01 &   8.63   &   29.93 &  8.31  &  30.80   &   8.43  &   30.48 \\
  $^{26}$Si+$^{26}$Si    &   0.857  &   8.83   &   29.57 &   8.77   &   29.51 &  8.49  &  30.16   &   8.60  &   29.94 \\
  $^{26}$Si+$^{28}$Si    &   0.928  &   8.97   &   29.17 &   8.91   &   29.12 &  8.66  &  29.61   &   8.75  &   29.48 \\
  $^{28}$Si+$^{28}$Si    &   1      &   9.09   &   28.80 &   9.04   &   28.74 &  8.85  &  29.08   &   8.89  &   29.07 \\
  $^{28}$Si+$^{30}$Si    &   1.071  &   9.20   &   28.46 &   9.16   &   28.40 &  9.00  &  28.61   &   9.01  &   28.72 \\
  $^{30}$Si+$^{30}$Si    &   1.143  &   9.31   &   28.15 &   9.29   &   28.06 &  9.16  &  28.17   &   9.12  &   28.41 \\
  \\
  $^{26}$Si+$^{52}$Ni    &   0.857  &   9.49   &  55.12 &   9.42    &   55.07 &   9.18   &  56.28   &   9.30   &   55.81  \\
  $^{26}$Si+$^{54}$Ni    &   0.904  &   9.57   &  54.70 &   9.50    &   54.66 &   9.28   &  55.77   &   9.39   &   55.32  \\
  $^{26}$Si+$^{56}$Ni    &   0.952  &   9.65   &  54.29 &   9.58    &   54.26 &   9.36   &  55.29   &   9.47   &   54.87  \\
  $^{28}$Si+$^{52}$Ni    &   0.904  &   9.62   &  54.41 &   9.55    &   54.39 &   9.36   &  55.32   &   9.44   &   55.04  \\
  $^{28}$Si+$^{54}$Ni    &   0.952  &   9.70   &  54.03 &   9.64    &   53.98 &   9.45   &  54.83   &   9.53   &   54.60  \\
  $^{28}$Si+$^{58}$Ni    &   1.048  &   9.84   &  53.30 &   9.79    &   53.22 &   9.62   &  53.94   &   9.68   &   53.81  \\
  $^{28}$Si+$^{62}$Ni    &   1.143  &   9.98   &  52.63 &   9.94    &   52.52 &   9.78   &  53.14   &   9.80   &   53.16  \\
  $^{28}$Si+$^{64}$Ni    &   1.190  &   10.04  &  52.31 &   10.01   &   52.19 &   9.85   &  52.78   &   9.86   &   52.88  \\
  $^{30}$Si+$^{58}$Ni    &   1.095  &   9.96   &  52.73 &   9.92    &   52.62 &   9.78   &  53.14   &   9.79   &   53.22  \\
  $^{30}$Si+$^{62}$Ni    &   1.190  &   10.09  &  52.10 &   10.07   &   51.94 &   9.94   &  52.37   &   9.92   &   52.61  \\
  $^{30}$Si+$^{64}$Ni    &   1.238  &   10.15  &  51.80 &   10.14   &   51.61 &   10.01  &  52.01   &   9.97   &   52.35  \\
  \\
  $^{34}$Ca+$^{34}$Ca    &   0.7   &   9.09    &   58.45 &   9.02    &  57.29  &   8.69   &   60.26   &   8.83  &   59.69 \\
  $^{36}$Ca+$^{36}$Ca    &   0.8   &   9.33    &   57.10 &   9.25    &  57.05  &   8.99   &   58.52   &   9.12  &   57.96 \\
  $^{38}$Ca+$^{38}$Ca    &   0.9   &   9.54    &   55.94 &   9.47    &  55.92  &   9.26   &   57.01   &   9.36  &   56.60 \\
  $^{40}$Ca+$^{34}$Ca    &   0.85  &   9.44    &   56.50 &   9.35    &  56.53  &   9.10   &   57.87   &   9.24  &   57.30 \\
  $^{40}$Ca+$^{36}$Ca    &   0.9   &   9.54    &   55.94 &   9.47    &  55.94  &   9.25   &   57.06   &   9.36  &   56.62 \\
  $^{40}$Ca+$^{38}$Ca    &   0.95  &   9.64    &   55.42 &   9.57    &  55.39  &   9.38   &   56.33   &   9.47  &   56.02 \\
  $^{40}$Ca+$^{40}$Ca    &   1     &   9.74    &   54.92 &   9.68    &  54.88  &   9.50   &   55.67   &   9.57  &   55.48 \\
  $^{40}$Ca+$^{44}$Ca    &   1.1   &   9.91    &   54.01 &   9.87    &  53.93  &   9.72   &   54.51   &   9.75  &   54.55 \\
  $^{40}$Ca+$^{48}$Ca    &   1.2   &   10.08   &   53.18 &   10.05   &  53.08  &   9.92   &   53.51   &   9.90  &   53.78 \\
  $^{48}$Ca+$^{48}$Ca    &   1.4   &   10.38   &   51.75 &   10.42   &  51.40  &   10.33  &   51.52   &   10.18 &   52.40 \\
  \hline
\end{tabular}
\end{center}
(a) Based on the AW 95 potential (b) Based on the Bass 80 potential
\\
(c) Based on the Denisov DP potential (d) Based on the Prox. 2010
potential

\newpage
Table 1. (Continued)

\begin{center}
\begin{tabular}{c c c c c c c c c c}
  \hline
  \hline
  Reaction & N/Z& $R_B^a$   &  $V_B^a$   &  $R_B^b$   &  $V_B^b$ & $R_B^c$   &  $V_B^c$ & $R_B^d$   &  $V_B^d$  \\
  &  &  (fm) & (MeV)& (fm)& (MeV) & (fm) & (MeV) & (fm) & (MeV) \\
  \hline
  $^{38}$Ca+$^{38}$Ti    &   0.809   &   9.44    &   62.06 &   9.36    &   62.04 &   9.10   &   63.58 &   9.24   &   62.97 \\
  $^{38}$Ca+$^{40}$Ti    &   0.857   &   9.55    &   61.46 &   9.47    &   61.44 &   9.24   &   62.78 &   9.55   &   61.46 \\
  $^{38}$Ca+$^{42}$Ti    &   0.857   &   9.64    &   60.91 &   9.57    &   60.88 &   9.36   &   62.05 &   9.64   &   60.91 \\
  $^{38}$Ca+$^{44}$Ti    &   0.952   &   9.73    &   60.38 &   9.67    &   60.36 &   9.47   &   61.38 &   9.57   &   61.02 \\
  $^{40}$Ca+$^{38}$Ti    &   0.857   &   9.55    &   61.42 &   9.47    &   61.44 &   9.23   &   62.81 &   9.36   &   62.25 \\
  $^{40}$Ca+$^{40}$Ti    &   0.904   &   9.65    &   60.86 &   9.57    &   60.86 &   9.36   &   62.03 &   9.47   &   61.58 \\
  $^{40}$Ca+$^{42}$Ti    &   0.952   &   9.74    &   60.34 &   9.67    &   60.32 &   9.48   &   61.32 &   9.58   &   60.98 \\
  $^{40}$Ca+$^{46}$Ti    &   1.048   &   9.91    &   59.37 &   9.86    &   59.30 &   9.70   &   60.07 &   9.76   &   59.95 \\
  $^{40}$Ca+$^{48}$Ti    &   1.095   &   10.00   &   58.92 &   9.95    &   58.84 &   9.80   &   59.51 &   9.84   &   59.50 \\
  $^{40}$Ca+$^{50}$Ti    &   1.143   &   10.07   &   58.49 &   10.04   &   58.40 &   9.90   &   59.00 &   9.91   &   59.09 \\
  \\
  $^{26}$S+$^{52}$Ni     &   0.772   &   9.34    &   63.79 &   9.26    &   63.71 &   8.94   &   65.75 &   9.14   &   64.80 \\
  $^{26}$S+$^{56}$Ni     &   0.863   &   9.52    &   62.71 &   9.43    &   62.75 &   9.13   &   64.56 &   9.34   &   63.55 \\
  $^{28}$S+$^{52}$Ni     &   0.818   &   9.48    &   62.95 &   9.40    &   62.89 &   9.14   &   64.52 &   9.29   &   63.81 \\
  $^{28}$S+$^{56}$Ni     &   0.909   &   9.65    &   61.96 &   9.57    &   61.96 &   9.32   &   63.38 &   9.48   &   62.68 \\
  $^{30}$S+$^{52}$Ni     &   0.863   &   9.61    &   62.18 &   9.54    &   62.15 &   9.31   &   63.46 &   9.43   &   62.94 \\
  $^{30}$S+$^{56}$Ni     &   0.954   &   9.77    &   61.27 &   9.70    &   61.24 &   9.50   &   62.36 &   9.61   &   61.91 \\
  $^{32}$S+$^{58}$Ni     &   1.045   &   9.95    &   60.23 &   9.90    &   60.15 &   9.74   &   60.97 &   9.80   &   60.08 \\
  $^{32}$S+$^{64}$Ni     &   1.182   &   10.16   &   59.12 &   10.12   &   59.00 &   9.97   &   59.67 &   9.99   &   59.75 \\
  $^{34}$S+$^{58}$Ni     &   1.090   &   10.06   &   59.65 &   10.02   &   59.53 &   9.88   &   60.18 &   9.90   &   60.20 \\
  $^{34}$S+$^{64}$Ni     &   1.227   &   10.25   &   58.61 &   10.24   &   58.41 &   10.11  &   58.92 &   10.08  &   59.21 \\
  $^{36}$S+$^{58}$Ni     &   1.136   &   10.16   &   59.11 &   10.13   &   58.96 &   10.01  &   59.47 &   10.00  &   59.65 \\
  $^{36}$S+$^{64}$Ni     &   1.273   &   10.34   &   58.12 &   10.35   &   57.85 &   10.24  &   58.23 &   10.18  &   58.72 \\
  \\
  $^{34}$Ar+$^{52}$Ni    &   0.870   &   9.72    &   69.15 &   9.65    &   69.11 &   9.42   &   70.54 &   9.57   &   69.95 \\
  $^{34}$Ar+$^{54}$Ni    &   0.913   &   9.80    &   68.64 &   9.73    &   68.60 &   9.52   &   69.91 &   9.65   &   69.36 \\
  $^{34}$Ar+$^{56}$Ni    &   0.956   &   9.88    &   68.15 &   9.81    &   68.11 &   9.61   &   69.32 &   9.73   &   68.34 \\
  $^{36}$Ar+$^{52}$Ni    &   0.913   &   9.83    &   68.44 &   9.76    &   68.41 &   9.57   &   69.62 &   9.68   &   69.18 \\
  $^{36}$Ar+$^{54}$Ni    &   0.956   &   9.91    &   67.96 &   9.84    &   67.91 &   9.66   &   69.01 &   9.76   &   68.63 \\
  $^{40}$Ar+$^{58}$Ni    &   1.130   &   10.24   &   65.92 &   10.21   &   65.76 &   10.08  &   66.40 &   10.10  &   66.51 \\
  $^{40}$Ar+$^{60}$Ni    &   1.174   &   10.31   &   65.54 &   10.29   &   65.34 &   10.17  &   65.92 &   10.16  &   66.14 \\
  $^{40}$Ar+$^{62}$Ni    &   1.217   &   10.37   &   65.18 &   10.36   &   64.93 &   10.24  &   65.47 &   10.22  &   65.79 \\
  $^{40}$Ar+$^{64}$Ni    &   1.260   &   10.43   &   64.82 &   10.43   &   64.54 &   10.32  &   65.03 &   10.28  &   65.46 \\
  \hline
\end{tabular}
\end{center}
(a) Based on the AW 95 potential (b) Based on the Bass 80 potential
\\
(c) Based on the Denisov DP potential (d) Based on the Prox. 2010
potential

\newpage
Table 1. (Continued)

\begin{center}
\begin{tabular}{c c c c c c c c c c}
  \hline
  \hline
  Reaction & N/Z& $R_B^a$   &  $V_B^a$   &  $R_B^b$   &  $V_B^b$ & $R_B^c$   &  $V_B^c$ & $R_B^d$   &  $V_B^d$  \\
  &  &  (fm) & (MeV)& (fm)& (MeV) & (fm) & (MeV) & (fm) & (MeV) \\
  \hline
  $^{36}$Ca+$^{50}$Ni    &   0.791   &   9.63    &    77.40 &   9.55     &   77.33  &   9.28    &   79.27 &   9.46    &   78.47  \\
  $^{36}$Ca+$^{52}$Ni    &   0.833   &   9.72    &    76.78 &   9.64     &   76.73  &   9.39    &   78.51 &   9.56    &   77.72  \\
  $^{36}$Ca+$^{54}$Ni    &   0.875   &   9.80    &    76.18 &   9.72     &   76.16  &   9.49    &   77.81 &   9.65    &   77.04  \\
  $^{36}$Ca+$^{56}$Ni    &   0.916   &   9.88    &    75.62 &   9.80     &   75.61  &   9.58    &   77.15 &   9.74    &   76.42  \\
  $^{38}$Ca+$^{52}$Ni    &   0.875   &   9.83    &    76.02 &   9.75     &   75.98  &   9.53    &   77.52 &   9.68    &   76.86  \\
  $^{38}$Ca+$^{54}$Ni    &   0.916   &   9.91    &    75.46 &   9.83     &   75.42  &   9.63    &   76.83 &   9.77    &   76.23  \\
  $^{38}$Ca+$^{56}$Ni    &   0.958   &   9.98    &    74.93 &   9.91     &   75.89  &   9.72    &   76.20 &   9.85    &   75.65  \\
  $^{40}$Ca+$^{52}$Ni    &   0.916   &   9.93    &    75.31 &   9.85     &   75.28  &   9.66    &   76.61 &   9.79    &   76.08  \\
  $^{40}$Ca+$^{54}$Ni    &   0.958   &   10.01   &    74.78 &   9.94     &   74.73  &   9.75    &   75.95 &   9.87    &   75.50  \\
  $^{40}$Ca+$^{58}$Ni    &   1.042   &   10.15   &    73.80 &   10.10    &   73.70  &   9.93    &   74.74 &   10.02   &   74.45  \\
  $^{40}$Ca+$^{62}$Ni    &   1.125   &   10.29   &    72.90 &   10.25    &   72.76  &   10.09   &   73.67 &   10.15   &   73.56  \\
  \\
  $^{40}$Ti+$^{48}$Ni    &   0.760   &   9.65    &   84.93  &   9.56     &   84.84  &   9.28    &   87.07 &   9.47    &   86.20  \\
  $^{40}$Ti+$^{50}$Ni    &   0.8     &   9.74    &   84.23  &   9.65     &   84.16  &   9.39    &   86.20 &   9.58    &   85.32  \\
  $^{40}$Ti+$^{52}$Ni    &   0.84    &   9.82    &   83.56  &   9.74     &   83.51  &   9.50    &   85.38 &   9.68    &   84.52  \\
  $^{40}$Ti+$^{56}$Ni    &   0.92    &   9.99    &   82.32  &   9.91     &   82.30  &   9.69    &   83.92 &   9.86    &   83.14  \\
  $^{42}$Ti+$^{52}$Ni    &   0.88    &   9.92    &   82.80  &   9.84     &   82.76  &   9.62    &   84.41 &   9.79    &   83.67  \\
  $^{42}$Ti+$^{56}$Ni    &   0.96    &   10.08   &   81.64  &   10.01    &   81.58  &   9.82    &   82.98 &   9.98    &   82.37  \\
  $^{48}$Ti+$^{58}$Ni    &   1.12    &   10.4    &   79.30  &   10.37    &   79.30  &   10.23   &   80.00 &   10.29   &   79.95  \\
  $^{48}$Ti+$^{60}$Ni    &   1.16    &   10.47   &   78.84  &   10.45    &   78.84  &   10.31   &   79.43 &   10.35   &   79.49  \\
  $^{48}$Ti+$^{64}$Ni    &   1.24    &   10.60   &   77.97  &   10.59    &   77.97  &   10.47   &   78.37 &   10.47   &   78.67  \\
  $^{46}$Ti+$^{64}$Ni    &   1.2     &   10.52   &   78.49  &   10.50    &   78.49  &   10.37   &   79.06 &   10.39   &   79.19  \\
  $^{50}$Ti+$^{60}$Ni    &   1.2     &   10.55   &   78.32  &   10.54    &   78.32  &   10.41   &   78.77 &   10.42   &   78.98  \\
  \\
  $^{48}$Ni+$^{48}$Ni    &   0.714   &   9.72    &   106.92 &   9.63     &   106.69 &   9.30    &   109.78&   9.59    &   108.54 \\
  $^{50}$Ni+$^{50}$Ni    &   0.786   &   9.93    &   105.10 &   9.82     &   104.99 &   9.54    &   107.57&   9.81    &   106.32 \\
  $^{52}$Ni+$^{50}$Ni    &   0.821   &   10.01   &   104.26 &   9.91     &   104.19 &   9.65    &   106.56&   9.90    &   105.35 \\
  $^{54}$Ni+$^{50}$Ni    &   0.857   &   10.09   &   103.47 &   10.00    &   103.42 &   9.76    &   105.62&   10.00   &   104.46 \\
  $^{54}$Ni+$^{54}$Ni    &   0.928   &   10.26   &   102.00 &   10.18    &   101.91 &   9.97    &   103.76&   10.17   &   102.81 \\
  $^{54}$Ni+$^{56}$Ni    &   0.964   &   10.33   &   101.31 &   10.26    &   101.20 &   10.07   &   102.92&   10.25   &   102.07 \\
  $^{56}$Ni+$^{50}$Ni    &   0.892   &   10.18   &   102.71 &   10.09    &   102.69 &   9.86    &   104.74&   10.08   &   103.64 \\
  $^{56}$Ni+$^{52}$Ni    &   0.928   &   10.26   &   102.00 &   10.18    &   101.93 &   9.96    &   103.80&   10.17   &   102.83 \\
  $^{56}$Ni+$^{54}$Ni    &   0.964   &   10.33   &   101.31 &   10.26    &   101.20 &   10.07   &   102.92&   10.25   &   102.07 \\
  $^{58}$Ni+$^{58}$Ni    &   1.071   &   10.55   &   99.41  &   10.51    &   99.18  &   10.34   &   100.55&   10.47   &   100.08 \\
  $^{58}$Ni+$^{64}$Ni    &   1.178   &   10.75   &   97.71  &   10.74    &   97.36  &   10.59   &   98.52 &   10.66   &   98.42  \\
  $^{64}$Ni+$^{64}$Ni    &   1.286   &   10.94   &   96.17  &   10.96    &   95.61  &   10.83   &   96.57 &   10.84   &   96.92  \\
  \hline
\end{tabular}
\end{center}
(a) Based on the AW 95 potential (b) Based on the Bass 80 potential
\\
(c) Based on the Denisov DP potential (d) Based on the Prox. 2010
potential

\newpage
Table 2. The calculated values of constant coefficients $\alpha_i$
and $\alpha_i^{'}$ which are extracted for fitting to regular
linear-behavior of $R_B$, $V_C$, $V_N$ and $V_B$, Eqs. (16,17) and
(20,21), as a function of increasing neutron in both ranges $N/Z
\leq1$ and $N/Z\geq1$.

\begin{center}
\begin{tabular}{c c c c c c c c c}
  \hline
  \hline
  Proximity-model & $\alpha_1$ & $\alpha_2$ &  $\alpha_3$   &  $\alpha_4$   &  $\alpha^{'}_1$ & $\alpha^{'}_2$ &  $\alpha^{'}_3$   &  $\alpha^{'}_4$  \\
  \hline
  AW 95      &  21.24 & -20.88 & -23.13 & 51.29  &  16.18 & -14.31 & -16.06 & 31.45 \\
  Bass 80    &  21.86 & -20.43 & -17.14 & 57.68  &  18.26 & -15.33 &  -23.54 & 41.82\\
  Denisov DP   &  27.70 & -27.52 & -32.13 & 68.33  &  20.78 & -18.11 &  -19.98 & 38.56\\
  Prox. 2010 &  25.00 &  -24.30 & -27.02 & 55.00 & 16.57 & -13.63 & -14.90 & 28.80  \\
  \hline

\end{tabular}
\end{center}

Table 3. The calculated values of constant coefficients $\beta_i$
and $\beta_i^{'}$ which are extracted for fitting to regular
non-linear-behavior of $R_B$, $V_C$, $V_N$ and $V_B$, Eqs. (18) and
(23), as a function of increasing neutron in $0.5\leq N/Z \leq 1.6$
range.

\begin{center}
\begin{tabular}{c c c c c c c c c}
  \hline
  \hline
  Proximity-model & $\beta_1$ & $\beta_2$ &  $\beta_3$   &  $\beta_4$   &  $\beta^{'}_1$ & $\beta^{'}_2$ &  $\beta^{'}_3$   &  $\beta^{'}_4$  \\
  \hline
  AW 95      &  -11.25 & 19.38 & 11.03 & -17.55  &  21.75 & -18.08 & -28.69 & 39.86 \\
  Bass 80    &  -1.94 & 20.37 & 4.55 & -17.74  &  6.52 & -20.26 &  -14.90 & 49.67\\
  Denisov DP   &  -10.87 & 24.84 & 17.11 & -22.84 &  20.07 & -25.65 &  -35.30 & 52.69\\
  Prox. 2010 &  -15.84 &  20.35 & 23.01 & -19.31 & 26.59 & -21.30 & -41.85 & 41.88  \\
  \hline

\end{tabular}
\end{center}

Table 4. The calculated values of constant coefficients $\gamma$
which are extracted for fitting to regular linear-behavior of
$\sigma_{fus}$, Eq. (27), as a function of $N/Z-1$.

\begin{center}
\begin{tabular}{c c c}
  \hline
  \hline
  Proximity-model & $\gamma$ (for $E_{c.m.}=1.125V_B^0$) & $\gamma$ (for $E_{c.m.}=1.375V_B^0$)  \\
  \hline
  AW 95      &  172.21 & 80.94  \\
  Bass 80    &  178.82 & 87.60  \\
  Denisov DP   &  22.31  & 108.35 \\
  Prox. 2010 &  185.03 & 95.28  \\
  \hline

\end{tabular}
\end{center}

\newpage
\begin{figure}
\begin{center}
\includegraphics{Fig1.eps}
\end{center}
\vspace{15cm} \caption{}
\end{figure}

\newpage
\begin{figure}
\begin{center}
\includegraphics{Fig2.eps}
\end{center}
\vspace{16cm} \caption{}
\end{figure}

\newpage
\begin{figure}
\begin{center}
\includegraphics{Fig3.eps}
\end{center}
\vspace{14cm} \caption{}
\end{figure}

\newpage
\begin{figure}
\begin{center}
\includegraphics{Fig4.eps}
\end{center}
\vspace{14cm} \caption{}
\end{figure}

\newpage
\begin{figure}
\begin{center}
\includegraphics{Fig5.eps}
\end{center}
\vspace{14cm} \caption{}
\end{figure}

\newpage
\begin{figure}
\begin{center}
\includegraphics{Fig6.eps}
\end{center}
\vspace{16cm} \caption{}
\end{figure}

\newpage
\begin{figure}
\begin{center}
\includegraphics{Fig7.eps}
\end{center}
\vspace{16cm} \caption{}
\end{figure}

\end{document}